\documentclass[aps,prb,english,showpacs,twocolumn,superscriptaddress]{revtex4-2}
\usepackage{amsfonts}
\usepackage{amssymb}
\usepackage{amsmath}
\usepackage{graphicx}

\usepackage{epsfig}
\usepackage{subfigure}
\usepackage{color}
\usepackage{amsmath,bm}
\usepackage{booktabs}
\usepackage{appendix}
\usepackage[colorlinks=true, linkcolor=blue, citecolor=blue, urlcolor=blue]{hyperref}
\usepackage{orcidlink}
\begin{document}
	
	\title{Exact phase diagram of the XXZ Heisenberg chain in a staggered
		magnetic field}
	
	\author{H. P. Zhang\,\orcidlink{0009-0001-5289-4865}}

\affiliation{
	\href{https://gscaep.ac.cn/}
	{Graduate School of China Academy of Engineering Physics},
	Beijing 100193, China
}
	
	\author{Z. Song\,\orcidlink{0000-0002-3315-4589}}
	\email[Corresponding author: ]{songtc@nankai.edu.cn}
	\affiliation{
		\href{https://www.nankai.edu.cn/}
		{School of Physics, Nankai University},
		Tianjin 300071, China
	}
\begin{abstract}
We study the phase diagram of the spin-1/2 XXZ Heisenberg chain in a
staggered magnetic field. An exact solution is obtained along the phase
boundary, where the exchange anisotropy and the staggered-field strength
satisfy an exact analytical relation, yielding a highly degenerate manifold
of ground states with off-diagonal long-range order. We demonstrate
{that these exact ground states lie on the boundary between the ferromagnetic
phase and a gapless Luttinger liquid phase. Upon further decreasing the
anisotropy, the gapless phase eventually gives way to a gapped
quantum-disordered phase. Large-scale density-matrix renormalization group
calculations confirm this phase structure and indicate that the transition
across the exact boundary is of first order.} Our results provide an exact benchmark for understanding
quantum phase transitions driven by staggered magnetic fields in
one-dimensional quantum spin systems.
\end{abstract}

\maketitle

\section{Introduction}

\label{Introduction}

The spin-$1/2$ XXZ Heisenberg chain is one of the paradigmatic models in
quantum many-body physics and has played a central role in the study of
strongly correlated one-dimensional systems. Owing to its integrability via
the Bethe ansatz, the ground-state phase diagram is known exactly and
exhibits rich quantum critical behavior. The interplay between exchange
anisotropy and external perturbations, such as staggered magnetic fields,
bond alternation, or Dzyaloshinskii--Moriya interactions, can further
stabilize novel quantum phases and induce quantum phase transitions, making
the XXZ chain an ideal platform for investigating low-dimensional quantum
magnetism \cite{Affleck_1989,Giamarchi_2003}.

{Introducing a staggered magnetic field further enriches the ground-state
	phase diagram of the XXZ chain by modifying the critical
	Tomonaga--Luttinger liquid and opening a gap in part of the parameter space \cite{Alcaraz1995}.
	From the perspective of bosonization, the staggered field can act as a
	relevant perturbation to the gapless Luttinger liquid and drive the system
	into a massive phase described by the sine-Gordon field theory} \cite%
{Oshikawa_1997,Affleck_1999,Giamarchi_2003}. Depending on the exchange
anisotropy and the field strength, the competition between quantum
fluctuations and staggered spin polarization gives rise to various quantum
phases separated by continuous or first-order quantum phase transitions. In
recent years, considerable attention has been devoted to the interplay
between staggered magnetic fields and bond alternation, which can stabilize
topological phases and induce topological quantum phase transitions
characterized by Berry phases, polarization, and nonlocal order parameters
\cite{M_rquez_2024}.

Despite extensive studies of the XXZ chain in staggered magnetic fields \cite{Okamoto1996,Tsukano1998,Kuzmenko2009,Andraschko2014,Rutkevich2018}, an
exact determination of its phase diagram remains challenging because the
staggered field generally breaks the integrability of the model. Although
bosonization and numerical methods have established the emergence of gapped
phases and quantum critical behavior, the exact location of phase boundaries
and the nature of the corresponding ground states are known only in a few
special cases. Therefore, identifying exactly solvable regimes and
establishing exact phase boundaries are of fundamental importance for
understanding the quantum phase diagram of the XXZ chain in staggered
magnetic fields.

In this work, we investigate the ground-state phase diagram of the spin-$1/2$
XXZ Heisenberg chain in a staggered magnetic field. We identify a resonant
condition under which the exchange anisotropy and the staggered-field
strength satisfy an exact analytical relation. Remarkably, along this line,
the Hamiltonian possesses an underlying restricted spectrum-generating algebra (RSGA) \cite{Moudgalya2020,Hashimoto2026,Imai2025,Tang2022,LiuSun2026},
which enables the exact construction of a highly degenerate ground-state
manifold. These exact ground states exhibit off-diagonal long-range order (ODLRO) \cite{Penrose1956,Yang1962,Rensink1967,Girardeau1971,Yang1989}, with the ground-state degeneracy increasing linearly with the
system size. By combining the exact solution with density-matrix
renormalization group (DMRG) \cite{White1992,White1993,Oestlund1995,Schollwoeck2005,Schollwoeck2011} calculations, we establish the ground-state
phase diagram and demonstrate that {the exact ground-state manifold lies on
	the boundary separating the ferromagnetic phase from a gapless Luttinger
	liquid phase. At smaller anisotropy, the latter eventually gives way to a
	gapped quantum-disordered phase. We further show that the transition across
	the exact boundary is first order.} Our results uncover an unconventional mechanism for
the emergence of a highly degenerate ground-state manifold with ODLRO in a
one-dimensional quantum spin system and provide an exact benchmark for the
phase diagram and quantum criticality of the XXZ chain in a staggered
magnetic field.

The remainder of this paper is organized as follows. Section~\ref{Model
	Hamiltonian} introduces the model and discusses its phase structure. Section~%
\ref{Exact phase boundary} presents the exact solution on the resonant line
and the resulting highly degenerate ground-state manifold. Section~\ref%
{Quantum phase transition} analyzes the phase transition across the exact
phase boundary, while Sec.~\ref{DMRG results} presents DMRG results for the
phase diagram and provides numerical confirmation of the analytical results.
Finally, Sec.~\ref{Summary} summarizes our main findings.

\section{Model Hamiltonian}

\label{Model Hamiltonian}

We consider the spin-$1/2$ anisotropic XXZ Heisenberg chain in a staggered
magnetic field on a ring with $2N$ sites. The Hamiltonian is

\begin{eqnarray}
H &=&-J\sum_{j=1}^{2N}[s_{j}^{x}s_{j+1}^{x}+s_{j}^{y}s_{j+1}^{y}+\Delta
\left( s_{j}^{z}s_{j+1}^{z}-\frac{1}{4}\right) ]  \notag \\
&&-h\sum_{j=1}^{2N}(-1)^{j}s_{j}^{z},  \label{H}
\end{eqnarray}%
where $s_{j}^{\alpha }$ ($\alpha =x,y,z$) are spin-$1/2$ operators at site $%
j $, and periodic boundary conditions are imposed. Here $J$ is the
nearest-neighbor exchange coupling and $\Delta $ denotes the exchange
anisotropy. Throughout this work we set $J=1$ as the unit of energy. For
later convenience, the staggered field is parameterized as

\begin{equation}
h=\sinh q,
\end{equation}%
where $q$ is a real parameter. Specifically, under the transformation $%
U=\exp \left( -i\pi \sum_{j=1}^{2N}s_{j}^{y}\right) $, the Hamiltonian
satisfies the relation%
\begin{equation}
UH(h)U^{\dag }=H(-h).  \label{reflection symmetry}
\end{equation}%
The Hamiltonian possesses a combined parity--time ($PT$) symmetry, satisfying%
\begin{equation}
\left[ PT,H\right] =0,  \label{PT symmetry}
\end{equation}%
where $P$ denotes the spatial inversion operation exchanging the two
sublattices,%
\begin{equation}
P\mathbf{s}_{A}P^{-1}=\mathbf{s}_{B},
\end{equation}%
or equivalently,%
\begin{equation}
P\mathbf{s}_{2j}P^{-1}=\mathbf{s}_{2j+1},
\end{equation}%
for the present lattice convention. The time-reversal operator ($T$) is
defined by%
\begin{equation}
T\mathbf{s}_{j}T^{-1}=-\mathbf{s}_{j}.
\end{equation}

\begin{figure}[t]
	\centering
	\includegraphics[width=0.95\linewidth]{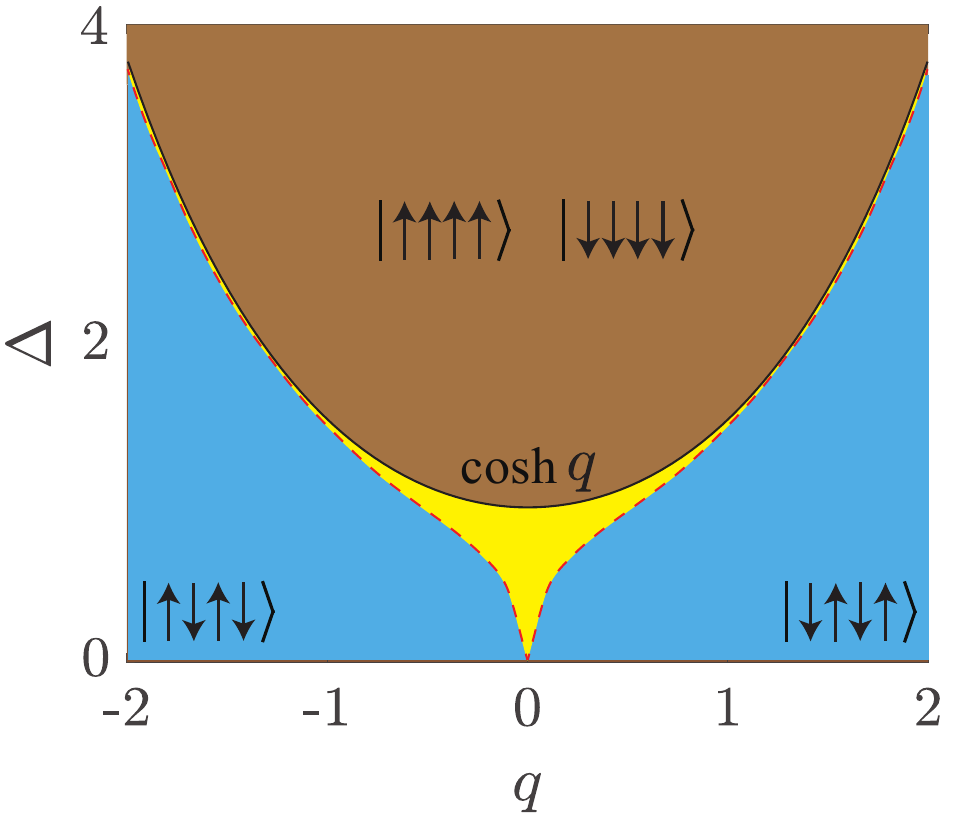}
\caption{Ground-state phase diagram of the Hamiltonian in Eq.~(\ref{H}) in the
$(q,\Delta)$ plane, with $h=\sinh q$ and $J=1$. The brown, yellow, and blue
regions denote the ferromagnetic, gapless Luttinger liquid, and gapped
quantum-disordered phases, respectively. The solid black curve gives the
exact boundary $\Delta=\cosh q$ between the ferromagnetic and gapless
phases, while the red dashed curve indicates the boundary between the
gapless and gapped nonferromagnetic phases. The gapless region narrows as
$|q|$ increases. The spin configurations schematically illustrate uniform
and staggered polarization. The curvature of the exact boundary shows that
a larger exchange anisotropy is required to stabilize the fully polarized
phase as the magnitude of the staggered field increases. The narrowing of
the yellow region reflects the increasing tendency of the staggered field
to gap the Luttinger liquid.}
	\label{fig:fig1}
\end{figure}

In the absence of the staggered field ($h=0$), the XXZ chain is exactly
solvable by the Bethe ansatz and possesses a well-established ground-state
phase diagram \cite{Bethe_1931,Yang_1966,Baxter_1985,Giamarchi_2003}. For $%
\Delta >1$, the ground state is a fully polarized ferromagnet separated from
excited states by a finite energy gap. In the easy-plane regime $-1\leq
\Delta <1$, the system realizes a gapless Tomonaga--Luttinger liquid
characterized by algebraically decaying spin correlations. For $\Delta <-1$,
a Berezinskii--Kosterlitz--Thouless transition drives the system into an
antiferromagnetic N\'{e}el phase with long-range staggered order and a
finite spin gap \cite{Luther_1975,Giamarchi_2003}.

When $\Delta =0$, the Hamiltonian in Eq.~(\ref{H}) can be mapped onto a
noninteracting spinless-fermion model through the Jordan--Wigner
transformation \cite{Jordan_1928,Lieb_1961}. The corresponding
single-particle spectrum is%
\begin{equation}
\varepsilon _{k}=\pm \sqrt{\cos ^{2}\frac{k}{2}+\sinh ^{2}q},
\end{equation}%
from which the ground-state energy is obtained by filling all
negative-energy states. The staggered magnetic field immediately opens a
finite excitation gap

\begin{equation}
\Delta _{\mathrm{g}}=2|\sinh q|,
\end{equation}%
indicating that the system evolves from a gapless Luttinger liquid to a
gapped quantum-disordered phase.

More generally, the asymptotic behavior of correlation functions provides a
fundamental criterion for distinguishing gapless and gapped one-dimensional
quantum phases. In a Tomonaga--Luttinger liquid, {the transverse spin
correlation function decays algebraically},%
\begin{equation}
C(r)=\langle s_{i}^{+}s_{i+r}^{-}\rangle \propto r^{-\eta },  \label{Cr}
\end{equation}%
where $\eta $ depends on the Luttinger parameter. {By contrast, in the
gapped quantum-disordered phase, the same correlation function decays
exponentially},

\begin{equation}
C(r)\propto e^{-r/\xi },  \label{correlation}
\end{equation}%
{where $\xi $ is the correlation length. A finite $\xi$ characterizes
short-range transverse correlations in this phase} \cite{Giamarchi_2003,Sachdev_1999,Hastings_2004}. Motivated
by the exactly solvable $\Delta =0$ limit, {it is natural to expect that the staggered magnetic field can generally drive the XXZ chain from the gapless Tomonaga–Luttinger liquid into a gapped quantum-disordered phase.}

In general, the Hamiltonian $H$ possesses a $U(1)$ symmetry associated with
the conservation of the total magnetization, 
\begin{equation}
s^{\alpha }=\sum_{j}s_{j}^{\alpha },\qquad (\alpha =x,y,z),
\end{equation}%
namely, 
\begin{equation}
\lbrack s^{z},H]=0.
\end{equation}%
This symmetry allows numerical simulations to be performed independently
within each invariant magnetization sector.

At the special point $\Delta =1$ and $h=0$, the Hamiltonian further
possesses the full $SU(2)$ spin-rotation symmetry, 
\begin{equation}
\lbrack s^{\alpha },H]=0,\qquad \alpha =x,y,z.
\end{equation}%
Consequently, the ground states form a spin multiplet with $(2N+1)$-fold
degeneracy. Starting from the fully polarized ferromagnetic state

\begin{figure*}[t]
	\centering
	\includegraphics[width=0.95\linewidth]{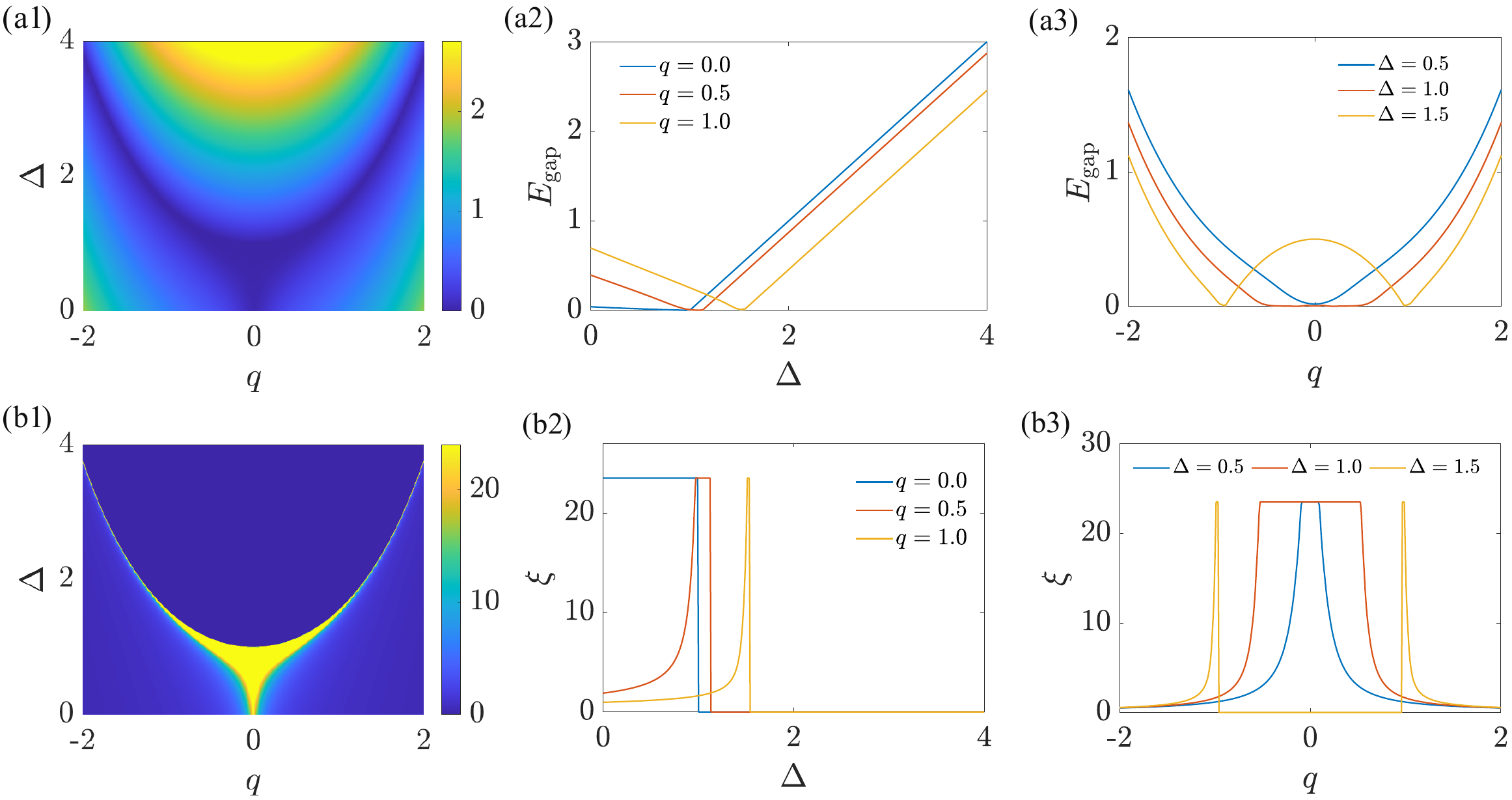}
\caption{Density-matrix renormalization group results for the excitation gap
$E_{\mathrm{gap}}$ (top row) and the fitted transverse correlation length
$\xi$ (bottom row), for $2N=40$ sites and $J=1$.
(a1), (b1) Color maps in the $(q,\Delta)$ plane.
(a2), (b2) Dependence on $\Delta$ at $q=0$, $0.5$, and $1.0$.
(a3), (b3) Dependence on $q$ at $\Delta=0.5$, $1.0$, and $1.5$.
The correlation length is extracted by fitting the transverse spin
correlations to the exponential form in Eq.~(\ref{correlation}); values
exceeding 24 lattice spacings are displayed as 24. Small finite-size gaps
and large fitted correlation lengths are consistent with the gapless
region shown in Fig.~\ref{fig:fig1}. The simultaneous suppression of the
gap and enhancement of $\xi$ therefore provide complementary finite-size
signatures of the gapless Luttinger-liquid region. As $|q|$ increases,
these signatures are confined to a narrower interval of $\Delta$, in
agreement with the shrinking gapless region in Fig.~\ref{fig:fig1}.}
	\label{fig:fig2}
\end{figure*}

\begin{equation}
|\Downarrow \rangle =\prod_{j}|\downarrow \rangle _{j},  \label{FM}
\end{equation}%
the degenerate ground-state manifold can be generated by applying the
spin-raising operator,

\begin{equation}
\frac{1}{n!\sqrt{C_{2N}^{n}}}(s^{+})^{n}|\Downarrow \rangle ,\qquad
s^{+}=s^{x}+is^{y},  \label{q=0}
\end{equation}%
with $n=0,1,\ldots ,2N$, based on the spectrum-generating algebra (SGA) \cite%
{Barut_1965,CABIBBO_2006}. The existence of this highly degenerate
ground-state manifold raises an intriguing question: whether a similar
degeneracy can survive in the presence of a staggered magnetic field, and if
such a degeneracy is associated with a critical phase boundary. Addressing
this question is the main objective of the present work.

\section{Exact phase boundary}

\label{Exact phase boundary}

In this section, we investigate the ground-state properties of the
Hamiltonian under a resonant condition, where the exchange anisotropy and
the staggered-field strength satisfy an exact analytical relation. We show
that, along this {phase-boundary line} in the ($q,\Delta $) parameter plane, the
Hamiltonian admits an exact solution with a highly degenerate manifold of
ground states. These exact ground states exhibit ODLRO and {lie
precisely on the phase boundary separating the ferromagnetic and gapless
Luttinger liquid phases}. As will be shown below, this remarkable degeneracy
originates from an underlying RSGA,
which enables the exact construction of the entire ground-state manifold and
provides a rigorous characterization of the {phase-boundary line}.

We consider the Hamiltonian under the condition%
\begin{equation}
\Delta =\Delta _{q}=\cosh q,
\end{equation}%
that is in the form

\begin{eqnarray}
H_{q} &=&-\sum_{j=1}^{2N}[s_{j}^{x}s_{j+1}^{x}+s_{j}^{y}s_{j+1}^{y}+\cosh
q\left( s_{j}^{z}s_{j+1}^{z}-\frac{1}{4}\right)  \notag \\
&&+\sinh q(-1)^{j}s_{j}^{z}].  \label{H_q}
\end{eqnarray}%
We note that the state $|\Downarrow \rangle $ is still the ground state of
the Hamiltonian with zero energy, i.e., $H_{q}\left\vert \Downarrow
\right\rangle =0$.

We introduce a set of pseudospin operators%
\begin{eqnarray}
L^{\pm } &=&\sum_{j\in \text{\textrm{odd}}}s_{j}^{\pm }+e^{\pm q}\sum_{j\in 
\text{\textrm{even}}}s_{j}^{\pm }, \\
L^{z} &=&\frac{1}{2}\left[ L^{+},L^{-}\right] =\sum_{j}s_{j}^{z},
\end{eqnarray}%
which obey the $SU(2)$ Lie algebra, i.e., $[L^{+},L^{-}]=2L^{z}$, and $%
[L^{z},L^{\pm }]=\pm L^{\pm }$. {For $q\neq0$, this is a
nonunitary realization because $L^{-}\neq(L^{+})^{\dagger}$.} Straightforward derivations show that%
\begin{eqnarray}
\left[ L^{+},\left[ L^{+},H_{q}\right] \right] &=&0, \\
\left[ L^{+},H_{q}\right] \left\vert \Downarrow \right\rangle &=&0,
\end{eqnarray}%
which meet the conditions of the RSGA. {The RSGA therefore generates
the following family of degenerate eigenstates of $H_{q}$:}%
\begin{equation}
\left\vert \psi _{q}^{n}\right\rangle =\frac{1}{n!\sqrt{{\Omega _n}}}%
(L^{+})^{n}\left\vert \Downarrow \right\rangle ,
\end{equation}%
with $n=0,1,\ldots ,2N$ and {$\Omega _n=\sum_{k=\max(0,n-N)}^{\min(n,N)}
e^{2kq}C_{N}^{n-k}C_{N}^{k}$}.
We note that $\left\vert \psi _{0}^{n}\right\rangle $\ reduces to the state
in Eq.~(\ref{q=0}), and $\left\vert \psi _{q}^{2N}\right\rangle =\left\vert
\Uparrow \right\rangle =\prod_{j}|\uparrow \rangle _{j}$.\ 

Indeed, by considering a superposition of the degenerate eigenstates $%
{\{|\psi _{q}^{n}\rangle \}}$ in the form

\begin{equation}
|\Phi (\theta )\rangle =\sum_{n}d_{n}{|\psi _{q}^{n}\rangle} ,
\end{equation}%
where

\begin{equation}
d_{n}=\frac{i^{n}\sqrt{{\Omega _n}}\sin ^{n}\left( \frac{\theta }{2}\right) \cos
^{2N-n}\left( \frac{\theta }{2}\right) }{\left[ 1+(e^{2q}-1)\sin ^{2}\left( 
\frac{\theta }{2}\right) \right] ^{N/2}},
\end{equation}%
we find that this state can be factorized into a direct product of
single-site states,

\begin{equation}
|\Phi (\theta )\rangle =\prod_{j=1}^{2N}|\phi _{j}(\theta )\rangle ,
\end{equation}%
where

\begin{equation}
|\phi _{j}(\theta )\rangle =%
\begin{cases}
\displaystyle\cos \left( \frac{\theta }{2}\right) |\downarrow \rangle
_{j}+i\sin \left( \frac{\theta }{2}\right) |\uparrow \rangle _{j}, & j\in 
\mathrm{odd}, \\[8pt] 
\displaystyle\frac{\cos \left( \frac{\theta }{2}\right) |\downarrow \rangle
_{j}+ie^{q}\sin \left( \frac{\theta }{2}\right) |\uparrow \rangle _{j}}{%
\sqrt{1+(e^{2q}-1)\sin ^{2}(\frac{\theta }{2})}}, & j\in \mathrm{even}.%
\end{cases}%
\end{equation}%
{Therefore, $|\Phi (\theta )\rangle $ is a tensor product state} composed of
locally polarized spins. This factorized form reveals that the degenerate
manifold contains symmetry-breaking states with long-range magnetic order.
Specifically, the correlation function $C(r)$ of the state $|\Phi (\theta
)\rangle $, defined in Eq.~(\ref{Cr}), remains finite and independent of $r$
at $\eta =0$, demonstrating the presence of ODLRO.

\begin{figure*}[tbp]
	\centering
	\includegraphics[width=0.95\linewidth]{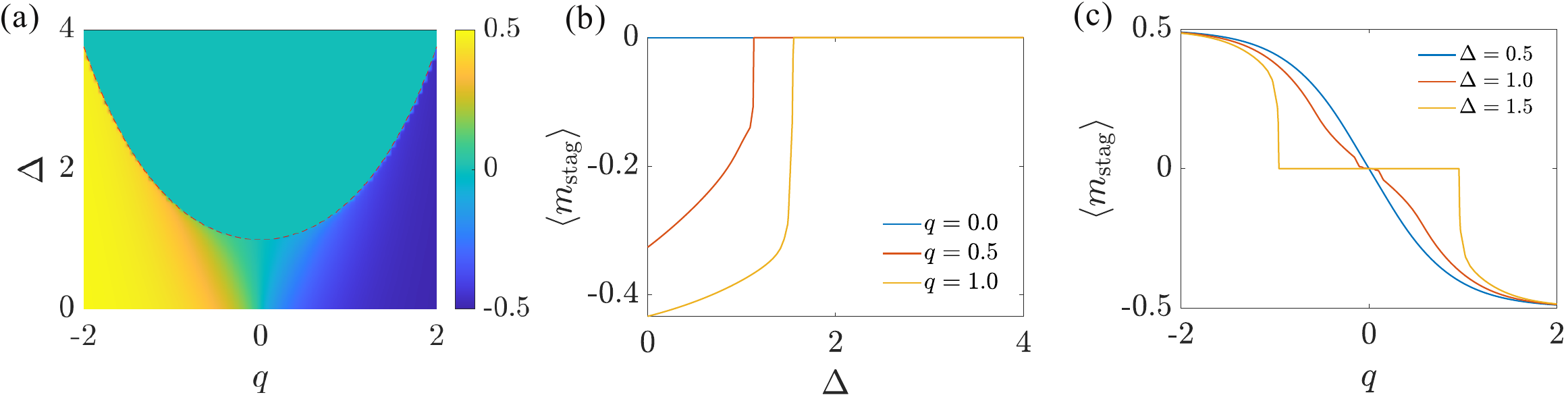}
\caption{Ground-state staggered magnetization $m_{\mathrm{stag}}$, defined in
Eq.~(\ref{m_stag}), for $2N=40$ sites and $J=1$.
(a) Color map in the $(q,\Delta)$ plane.
(b) Dependence on $\Delta$ at $q=0$, $0.5$, and $1.0$.
(c) Dependence on $q$ at $\Delta=0.5$, $1.0$, and $1.5$.
The staggered magnetization vanishes in the fully polarized ferromagnetic
region, $\Delta>\cosh q$, and reverses sign under $q\to -q$. The vanishing
in the ferromagnetic region follows because a uniform polarization cancels
in the staggered sum, whereas finite values below the exact boundary
reflect the sublattice polarization induced by the staggered field. The
odd response under $q\to -q$ reflects the symmetry relating opposite
staggered fields, and the approach to zero at $\Delta=\cosh q$ tracks the
loss of staggered polarization on entering the ferromagnetic region.}
	\label{fig:fig3}
\end{figure*}

\section{Quantum phase transition}

\label{Quantum phase transition}

In this section, we investigate {the ferromagnetic and gapless Luttinger
liquid phases separated by the phase boundary $\Delta _{q}=\cosh q$} using analytical approaches and numerical
calculations. We first employ degenerate perturbation theory to analyze the
low-energy states. {For a finite system, this expansion is expected to
be most reliable sufficiently close to the exact boundary and provides
qualitative insight into the quantum phase transition.}

We consider the perturbation Hamiltonian

\begin{equation}
H_{\text{\textrm{pert}}}=H_{q}+H^{\prime },  \label{H_pert}
\end{equation}%
where the perturbation is given by an Ising interaction,%
\begin{equation}
H^{\prime }=-\lambda \sum_{j=1}^{2N}\left( s_{j}^{z}s_{j+1}^{z}-\frac{1}{4}%
\right) .
\end{equation}%
Within the degenerate manifold spanned by the states $\left\{ \left\vert
\psi _{q}^{n}\right\rangle \right\} $, the perturbation is diagonal, with
the nonzero matrix elements given by

\begin{equation}
\left\langle \psi _{q}^{m}\right\vert H^{\prime }\left\vert \psi
_{q}^{n}\right\rangle =-\lambda \delta _{mn}\sum_{j=1}^{2N}\left\langle \psi
_{q}^{n}\right\vert \left( s_{j}^{z}s_{j+1}^{z}-\frac{1}{4}\right)
\left\vert \psi _{q}^{n}\right\rangle .
\end{equation}%
Accordingly, the first-order energy correction for the state $\left\vert
\psi _{q}^{n}\right\rangle $\ is

\begin{eqnarray}
E(N,q,n) &=&\left\langle \psi _{q}^{n}\right\vert H^{\prime }\left\vert \psi
_{q}^{n}\right\rangle  \label{E1} \\
&=&\lambda \left( 1+e^{2q}\right) \frac{%
N\sum_{k=0}^{n-1}e^{2kq}C_{N-1}^{n-k-1}C_{N-1}^{k}}{%
\sum_{k=0}^{n}e^{2kq}C_{N}^{n-k}C_{N}^{k}},  \notag 
\end{eqnarray}%
for $n=1,2,\cdots ,N$, while $E(N,q,0)=0$.\ The energy corrections satisfy
the reflection relation $E(N,q,2N-n)=E(N,q,n)$. We next use these energy
corrections to determine the energetically favored states and characterize
the resulting quantum phases.

(i) In the region $\Delta >\Delta _{q}$ $\left( \lambda >0\right) $, the
degenerate ground states are $\left\vert \psi _{q}^{0}\right\rangle $\ and $%
\left\vert \psi _{q}^{2N}\right\rangle $. In fact, this result can be
established exactly. The original Hamiltonian can be decomposed into a sum
of local bond Hamiltonians,%
\begin{equation}
H=\sum_{j=1}^{2N}H_{j}
\end{equation}%
with%
\begin{eqnarray}
H_{j} &=&-J[s_{j}^{x}s_{j+1}^{x}+s_{j}^{y}s_{j+1}^{y}+\Delta \left(
s_{j}^{z}s_{j+1}^{z}-\frac{1}{4}\right) ]  \notag \\
&&{-\frac{h}{2}\left[(-1)^{j}s_{j}^{z}
+(-1)^{j+1}s_{j+1}^{z}\right]}.
\end{eqnarray}%
For the fully polarized states $\left\vert \Uparrow \right\rangle $\ and $%
\left\vert \Downarrow \right\rangle $, one readily finds%
\begin{equation}
H_{j}\left\vert \Uparrow \right\rangle =H_{j}\left\vert \Downarrow
\right\rangle =0.
\end{equation}%
Moreover, for $\Delta >\Delta _{q}$, these states minimize the local
Hamiltonian $H_{j}$. Consequently, the fully polarized states are exact
ground states of the full Hamiltonian. Since they correspond to $\left\vert
\psi _{q}^{0}\right\rangle $\ and $\left\vert \psi _{q}^{2N}\right\rangle $
respectively, {these two states form an exactly degenerate ground-state
doublet.} Furthermore, this analysis indicates that
the first excited state lies an energy $\lambda $ above the ground-state
manifold, yielding the excitation gap%
\begin{equation}
E_{\text{\textrm{gap}}}=\lambda .
\end{equation}%
As shown below, this exact result is in excellent agreement with the
numerical calculations.

(ii) {For $\Delta <\Delta _{q}$ $\left( \lambda <0\right) $, within
first-order degenerate perturbation theory and sufficiently close to the
exact boundary, the lowest-energy state in the projected manifold is
$\left\vert \psi _{q}^{N}\right\rangle $, and the predicted finite-size gap
to the first excited state is}%
\begin{equation}
E_{\text{\textrm{gap}}}=-\lambda \left[ \frac{\cosh q}{2N}+O\left( \frac{1}{%
N^{2}}\right) \right] .
\end{equation}%
{Thus, the perturbative gap vanishes in the thermodynamic limit, consistent
with the gapless phase immediately below the exact boundary.} In contrast to
the case of $\Delta >\Delta _{q}$, where the
ground states can be established exactly for finite $\lambda $, we cannot
analytically establish that $\left\vert \psi _{q}^{N}\right\rangle $ remains
an exact ground-state eigenstate for finite $\lambda $ in the present
regime. We therefore perform numerical calculations in the next section to
verify and complement the analytical results, providing a more complete
characterization of the quantum phases.

\begin{figure*}[tbp]
	\centering
	\includegraphics[width=0.9\linewidth]{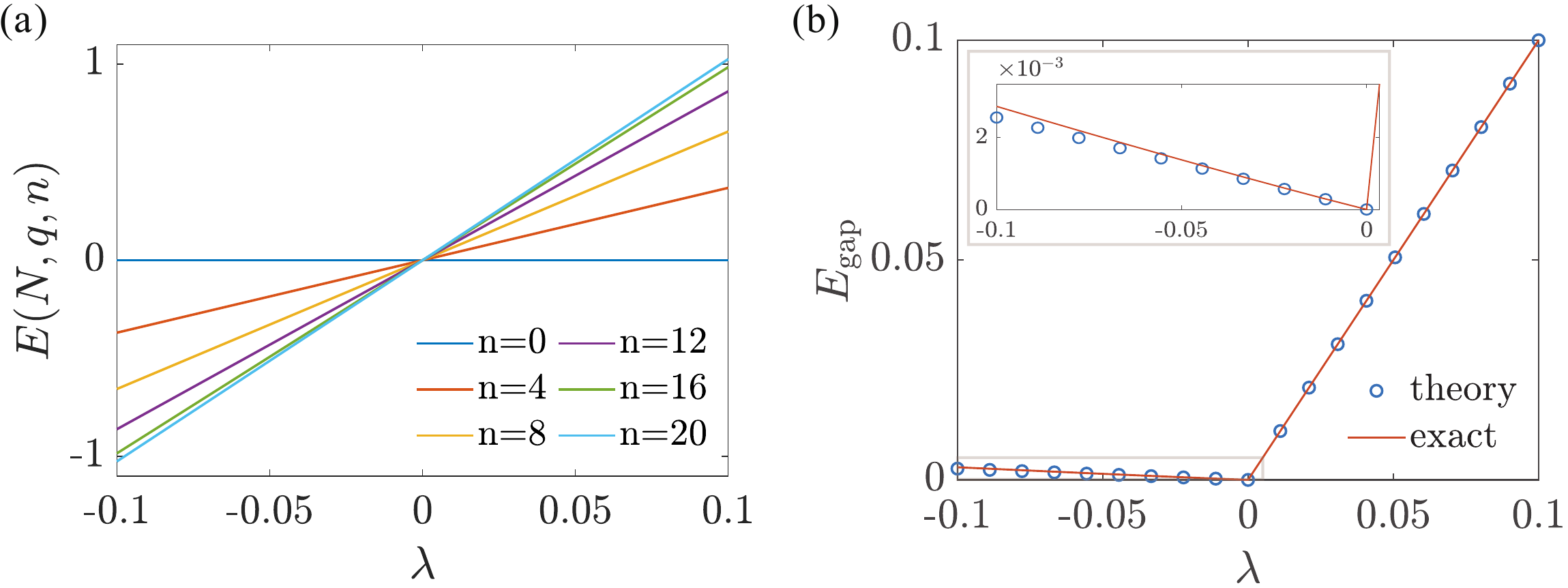}
\caption{First-order splitting of the degenerate ground-state manifold and the
excitation gap near $\Delta=\cosh q$.
(a) First-order energy corrections $E(N,q,n)$ from Eq.~(\ref{E1}), plotted
against $\lambda=\Delta-\cosh q$ for the selected sectors $n=0,4,\ldots,20$.
The $n=0$ level is degenerate with the fully polarized state at $n=2N$.
(b) Excitation gap $E_{\mathrm{gap}}$ as a function of $\lambda$.
Blue open circles denote the first-order perturbative prediction, and the
red solid line denotes the numerical result. The inset enlarges the
small-gap region for $\lambda<0$. Both panels use $2N=40$, $q=0$, and $J=1$.
For $\lambda>0$, the lowest levels are the endpoint states $n=0$ and
$n=2N$, corresponding to the fully polarized ground-state doublet,
whereas for $\lambda<0$ the lowest state moves toward the central sector
$n=N$. The change in the excitation-gap behavior across $\lambda=0$
therefore reflects the different low-energy selections on the two sides of
the exact boundary: the ferromagnetic-side gap is linear in $\lambda$,
while the perturbative gap on the $\lambda<0$ side decreases as $1/N$.}
	\label{fig:fig4}
\end{figure*}
\section{DMRG results}

\label{DMRG results}

In this section, we perform DMRG
calculations to verify and further examine our analytical results for
finite-size systems. We focus on three quantities that characterize the
phase structure and, in principle, can also be measured experimentally. Our
analytical results lead to the following predictions.

(i) \textit{Energy gap.} {The energy gap between the ground state and the
first excited state is expected to vanish in the thermodynamic limit
throughout the gapless Luttinger liquid region, including at the exact phase
boundary $\Delta=\cosh q$.} In addition, at $q=0$, the model
reduces to the zero-field XXZ chain, which is gapless in the region $%
-1<\Delta<1$.

(ii) \textit{Correlation length.} {The correlation length, defined in Eq.~(%
\ref{correlation}), is expected to become large or divergent throughout the
gapless Luttinger liquid region. By contrast, a finite correlation length
characterizes the short-range correlations in the gapped
quantum-disordered phase.}

(iii) \textit{Staggered magnetization.} We define the staggered
magnetization as 
\begin{equation}
m_{\mathrm{stag}}=\frac{1}{2N}\sum_{j=1}^{2N}(-1)^{j+1}\langle
s_{j}^{z}\rangle .
\label{m_stag}
\end{equation}%
As illustrated in Fig.~\ref{fig:fig1}, $m_{\mathrm{stag}}$ is expected to
vanish in the {ferromagnetic phase} for $\Delta >\cosh q$. In contrast, {it is
finite over the displayed range $0<\Delta <\cosh q$} and changes sign upon
crossing $q=0$. The results are presented in Figs.~\ref{fig:fig2} and \ref%
{fig:fig3}, with the other system parameters specified in the corresponding
captions. These results reveal the following features. (i) {The energy gap
approaches zero near the phase boundary and remains consistent with gapless
behavior in the Luttinger liquid region. (ii) The fitted correlation length
becomes large in the region identified as the gapless Luttinger liquid,
whereas it remains finite in the gapped quantum-disordered phase.} (iii) The staggered magnetization $m_{%
\mathrm{stag}}$ vanishes within the region $\Delta>\cosh q$. In contrast, it
{is finite over the displayed range $0<\Delta<\cosh q$} and changes sign upon
crossing $q=0$.

In addition, we investigate the validity of the perturbative approach
presented in the preceding section. Specifically, we calculate the energy
gap of the perturbed Hamiltonian $H_{\mathrm{pert}}$ for finite-size systems
as a function of $\lambda$. The results are shown in Fig.~\ref{fig:fig4},
where we compare the numerical results with the perturbative prediction $E_{%
\mathrm{gap}}=\lambda$. The numerical results indicate that the perturbative
approach remains accurate even for a finite perturbation strength of $%
\lambda=0.1$. We expect the range of validity of the perturbative
description to become narrower as the system size $N$ increases.
Nevertheless, the perturbative approach provides a clear physical picture of
the phase transition.

\bigskip 

\section{Summary}

\label{Summary}

In summary, we have investigated the ground-state phase diagram of the spin-$%
1/2$ XXZ Heisenberg chain in a staggered magnetic field. By parameterizing
the staggered field as $h=\sinh q$, we identify an exact resonant condition, 
$\Delta =\cosh q$,\ along which the Hamiltonian acquires an underlying
RSGA. This algebra allows us to construct an exact
family of ground states, forming a highly degenerate manifold whose
degeneracy increases linearly with the system size. The resulting states
possess ODLRO and provide an exact characterization
of the ground-state manifold on the phase boundary. Combining the exact
solution with DMRG calculations, we
establish the phase structure of the model and locate the exact phase
boundary associated with the resonant condition. {The exact ground-state
manifold lies at the boundary between the ferromagnetic and gapless
Luttinger liquid phases; at smaller anisotropy, the gapless phase eventually
gives way to a gapped quantum-disordered phase. Our analysis indicates that the
transition across the exact boundary is first order.} Our results demonstrate that a staggered magnetic field
can generate an exactly solvable, highly degenerate ground-state manifold in
an interacting one-dimensional spin system, despite the fact that the
staggered field generally destroys the integrability of the XXZ chain. More
broadly, our work illustrates how an emergent algebraic structure can lead
to exact many-body ground states and unconventional long-range order at a
phase boundary.

\section*{Acknowledgment}

This work was supported by the National Natural Science Foundation of China
(under Grant No.~12374461).

\section*{DATA AVAILABILITY}

The data that support the findings of this article are not publicly
available. The data are available from the authors upon reasonable request.

\bibliography{XXZ_QSL_Ref}

@incollection{Baxter_1985,
  author     = {Baxter, R. J.},
  title      = {Exactly solved models in statistical mechanics},
  booktitle  = {Integrable Systems in Statistical Mechanics},
  publisher  = {World Scientific},
  pages      = {5--63},
  year       = {1985},
  month      = may,
  doi        = {10.1142/9789814415255_0002},
  isbn       = {9789814415255},
  issn       = {2010-1996},
}

@article{Sachdev_1999,
  author     = {Sachdev, Subir},
  title      = {Quantum phase transitions},
  journal    = {Phys. World},
  publisher  = {IOP Publishing},
  volume     = {12},
  number     = {4},
  pages      = {33--38},
  year       = {1999},
  month      = apr,
  doi        = {10.1088/2058-7058/12/4/23},
  issn       = {2058-7058},
}

@incollection{Giamarchi_2003,
  author     = {Giamarchi, Thierry},
  title      = {Disordered systems},
  booktitle  = {Quantum Physics in One Dimension},
  publisher  = {Oxford University Press},
  chapter    = {9},
  pages      = {270--302},
  year       = {2003},
  month      = dec,
  doi        = {10.1093/acprof:oso/9780198525004.003.0009},
  isbn       = {9780198525004},
}

@article{Luther_1975,
  author     = {Luther, A. and Peschel, I.},
  title      = {Calculation of critical exponents in two dimensions from quantum field theory in one dimension},
  journal    = {Phys. Rev. B},
  publisher  = {American Physical Society (APS)},
  volume     = {12},
  number     = {9},
  pages      = {3908--3917},
  year       = {1975},
  month      = nov,
  doi        = {10.1103/PhysRevB.12.3908},
  issn       = {0556-2805},
}

@article{Jordan_1928,
  author     = {Jordan, P. and Wigner, E.},
  title      = {{{\"U}ber das Paulische {\"A}quivalenzverbot}},
  journal    = {Z. Phys.},
  publisher  = {Springer Science and Business Media LLC},
  volume     = {47},
  number     = {9--10},
  pages      = {631--651},
  year       = {1928},
  month      = sep,
  doi        = {10.1007/BF01331938},
  issn       = {1434-601X},
}

@article{Lieb_1961,
  author     = {Lieb, Elliott and Schultz, Theodore and Mattis, Daniel},
  title      = {Two soluble models of an antiferromagnetic chain},
  journal    = {Ann. Phys.},
  publisher  = {Elsevier BV},
  volume     = {16},
  number     = {3},
  pages      = {407--466},
  year       = {1961},
  month      = dec,
  doi        = {10.1016/0003-4916(61)90115-4},
  issn       = {0003-4916},
}

@article{Hastings_2004,
  author     = {Hastings, M. B.},
  title      = {{Lieb-Schultz-Mattis} in higher dimensions},
  journal    = {Phys. Rev. B},
  publisher  = {American Physical Society (APS)},
  volume     = {69},
  number     = {10},
  pages      = {104431},
  year       = {2004},
  month      = mar,
  doi        = {10.1103/PhysRevB.69.104431},
  issn       = {1550-235X},
}

@article{Bethe_1931,
  author     = {Bethe, H.},
  title      = {{Zur Theorie der Metalle: I. Eigenwerte und Eigenfunktionen der linearen Atomkette}},
  journal    = {Z. Phys.},
  publisher  = {Springer Science and Business Media LLC},
  volume     = {71},
  number     = {3--4},
  pages      = {205--226},
  year       = {1931},
  month      = mar,
  doi        = {10.1007/BF01341708},
  issn       = {1434-601X},
}

@article{Yang_1966,
  author     = {Yang, C. N. and Yang, C. P.},
  title      = {One-dimensional chain of anisotropic spin-spin interactions. {I}. {Proof} of {Bethe}'s hypothesis for ground state in a finite system},
  journal    = {Phys. Rev.},
  publisher  = {American Physical Society (APS)},
  volume     = {150},
  number     = {1},
  pages      = {321--327},
  year       = {1966},
  month      = oct,
  doi        = {10.1103/PhysRev.150.321},
  issn       = {0031-899X},
}

@article{Affleck_1989,
  author     = {Affleck, I.},
  title      = {Quantum spin chains and the {Haldane} gap},
  journal    = {J. Phys.: Condens. Matter},
  publisher  = {IOP Publishing},
  volume     = {1},
  number     = {19},
  pages      = {3047--3072},
  year       = {1989},
  month      = may,
  doi        = {10.1088/0953-8984/1/19/001},
  issn       = {1361-648X},
}

@article{Oshikawa_1997,
  author     = {Oshikawa, Masaki and Affleck, Ian},
  title      = {Field-induced gap in {$S=1/2$} antiferromagnetic chains},
  journal    = {Phys. Rev. Lett.},
  publisher  = {American Physical Society (APS)},
  volume     = {79},
  number     = {15},
  pages      = {2883--2886},
  year       = {1997},
  month      = oct,
  doi        = {10.1103/PhysRevLett.79.2883},
  issn       = {1079-7114},
}

@article{Affleck_1999,
  author     = {Affleck, Ian and Oshikawa, Masaki},
  title      = {Field-induced gap in {Cu} benzoate and other {$S=1/2$} antiferromagnetic chains},
  journal    = {Phys. Rev. B},
  publisher  = {American Physical Society (APS)},
  volume     = {60},
  number     = {2},
  pages      = {1038--1056},
  year       = {1999},
  month      = jul,
  doi        = {10.1103/PhysRevB.60.1038},
  issn       = {1095-3795},
}

@article{M_rquez_2024,
  author     = {M{\'a}rquez, B. F. and Aucar Boidi, N. and Hallberg, K. and Aligia, A. A.},
  title      = {Phase diagram and topology of the {XXZ} chain with alternating bonds and staggered magnetic field},
  journal    = {Phys. Rev. B},
  publisher  = {American Physical Society (APS)},
  volume     = {109},
  number     = {23},
  pages      = {235143},
  year       = {2024},
  month      = jun,
  doi        = {10.1103/PhysRevB.109.235143},
  issn       = {2469-9969},
}

@article{Barut_1965,
  author     = {Barut, A. O. and B{\"o}hm, A.},
  title      = {Dynamical groups and mass formula},
  journal    = {Phys. Rev.},
  publisher  = {American Physical Society (APS)},
  volume     = {139},
  number     = {4B},
  pages      = {B1107--B1112},
  year       = {1965},
  month      = aug,
  doi        = {10.1103/PhysRev.139.B1107},
  issn       = {0031-899X},
}

@incollection{CABIBBO_2006,
  author     = {Cabibbo, Nicola},
  title      = {Algebraic theory of particle physics and spectrum generating algebras},
  booktitle  = {Matter Particled---Patterns, Structure and Dynamics},
  publisher  = {World Scientific and China Waterpower Press},
  pages      = {65--167},
  year       = {2006},
  month      = mar,
  doi        = {10.1142/9789812774033_0003},
  issn       = {1793-1207},
}

@article{Alcaraz1995,
  author     = {Alcaraz, F. C. and Malvezzi, A. L.},
  title      = {Critical and off-critical properties of the {XXZ} chain in external homogeneous and staggered magnetic fields},
  journal    = {J. Phys. A: Math. Gen.},
  publisher  = {IOP Publishing},
  volume     = {28},
  number     = {6},
  pages      = {1521--1534},
  year       = {1995},
  month      = mar,
  doi        = {10.1088/0305-4470/28/6/009},
  issn       = {1361-6447},
}

@article{Moudgalya2020,
  author     = {Moudgalya, Sanjay and Regnault, Nicolas and Bernevig, B. Andrei},
  title      = {{$\eta$}-pairing in {Hubbard} models: From spectrum generating algebras to quantum many-body scars},
  journal    = {Phys. Rev. B},
  publisher  = {American Physical Society (APS)},
  volume     = {102},
  number     = {8},
  pages      = {085140},
  year       = {2020},
  month      = aug,
  doi        = {10.1103/PhysRevB.102.085140},
  issn       = {2469-9969},
}

@article{Imai2025,
  author     = {Imai, Shohei and Tsuji, Naoto},
  title      = {Quantum many-body scars with unconventional superconducting pairing symmetries via multibody interactions},
  journal    = {Phys. Rev. Research},
  publisher  = {American Physical Society (APS)},
  volume     = {7},
  number     = {1},
  pages      = {013064},
  year       = {2025},
  month      = jan,
  doi        = {10.1103/PhysRevResearch.7.013064},
  issn       = {2643-1564},
}

@article{LiuSun2026,
  author     = {Liu-Sun, J. Y. and Song, Z.},
  title      = {Resonant fields inducing energy towers in {Lieb} quantum spin lattices},
  journal    = {Phys. Rev. B},
  publisher  = {American Physical Society (APS)},
  volume     = {113},
  number     = {15},
  pages      = {155122},
  year       = {2026},
  month      = apr,
  doi        = {10.1103/m2xx-pg59},
  issn       = {2469-9969},
}

@article{Tang2022,
  author     = {Tang, Long-Hin and O'Dea, Nicholas and Chandran, Anushya},
  title      = {Multimagnon quantum many-body scars from tensor operators},
  journal    = {Phys. Rev. Research},
  publisher  = {American Physical Society (APS)},
  volume     = {4},
  number     = {4},
  pages      = {043006},
  year       = {2022},
  month      = oct,
  doi        = {10.1103/PhysRevResearch.4.043006},
  issn       = {2643-1564},
}

@article{Hashimoto2026,
  author     = {Hashimoto, Daiki and Kunimi, Masaya and Nikuni, Tetsuro},
  title      = {Construction of asymptotic quantum many-body scar states in the {$\mathrm{SU}(N)$} {Hubbard} model},
  journal    = {Phys. Rev. B},
  publisher  = {American Physical Society (APS)},
  volume     = {113},
  number     = {15},
  pages      = {155137},
  year       = {2026},
  month      = apr,
  doi        = {10.1103/35d7-qgx4},
  issn       = {2469-9969},
}

@article{Penrose1956,
  author     = {Penrose, Oliver and Onsager, Lars},
  title      = {{Bose-Einstein} condensation and liquid helium},
  journal    = {Phys. Rev.},
  publisher  = {American Physical Society (APS)},
  volume     = {104},
  number     = {3},
  pages      = {576--584},
  year       = {1956},
  month      = nov,
  doi        = {10.1103/PhysRev.104.576},
  issn       = {0031-899X},
}

@article{Yang1962,
  author     = {Yang, C. N.},
  title      = {Concept of off-diagonal long-range order and the quantum phases of liquid {He} and of superconductors},
  journal    = {Rev. Mod. Phys.},
  publisher  = {American Physical Society (APS)},
  volume     = {34},
  number     = {4},
  pages      = {694--704},
  year       = {1962},
  month      = oct,
  doi        = {10.1103/RevModPhys.34.694},
  issn       = {0034-6861},
}

@article{Rensink1967,
  author     = {Rensink, Marvin E.},
  title      = {Off-diagonal long-range order in the {BCS} theory},
  journal    = {Ann. Phys.},
  publisher  = {Elsevier BV},
  volume     = {44},
  number     = {1},
  pages      = {105--111},
  year       = {1967},
  month      = aug,
  doi        = {10.1016/0003-4916(67)90267-9},
  issn       = {0003-4916},
}

@article{Girardeau1971,
  author     = {Girardeau, M. D.},
  title      = {Off-diagonal long-range order and the momentum distribution of electron pairs in superconductors, and helium atoms in liquid {$^{4}\mathrm{He}$}},
  journal    = {Phys. Rev. A},
  publisher  = {American Physical Society (APS)},
  volume     = {4},
  number     = {2},
  pages      = {777--788},
  year       = {1971},
  month      = aug,
  doi        = {10.1103/PhysRevA.4.777},
  issn       = {0556-2791},
}

@article{Yang1989,
  author     = {Yang, Chen Ning},
  title      = {{$\eta$}-pairing and off-diagonal long-range order in a {Hubbard} model},
  journal    = {Phys. Rev. Lett.},
  publisher  = {American Physical Society (APS)},
  volume     = {63},
  number     = {19},
  pages      = {2144--2147},
  year       = {1989},
  month      = nov,
  doi        = {10.1103/PhysRevLett.63.2144},
  issn       = {0031-9007},
}

@article{White1992,
  author     = {White, Steven R.},
  title      = {Density matrix formulation for quantum renormalization groups},
  journal    = {Phys. Rev. Lett.},
  publisher  = {American Physical Society (APS)},
  volume     = {69},
  number     = {19},
  pages      = {2863--2866},
  year       = {1992},
  month      = nov,
  doi        = {10.1103/PhysRevLett.69.2863},
  issn       = {0031-9007},
}

@article{White1993,
  author     = {White, Steven R.},
  title      = {Density-matrix algorithms for quantum renormalization groups},
  journal    = {Phys. Rev. B},
  publisher  = {American Physical Society (APS)},
  volume     = {48},
  number     = {14},
  pages      = {10345--10356},
  year       = {1993},
  month      = oct,
  doi        = {10.1103/PhysRevB.48.10345},
  issn       = {1095-3795},
}

@article{Oestlund1995,
  author     = {{\"O}stlund, Stellan and Rommer, Stefan},
  title      = {Thermodynamic limit of density matrix renormalization},
  journal    = {Phys. Rev. Lett.},
  publisher  = {American Physical Society (APS)},
  volume     = {75},
  number     = {19},
  pages      = {3537--3540},
  year       = {1995},
  month      = nov,
  doi        = {10.1103/PhysRevLett.75.3537},
  issn       = {1079-7114},
}

@article{Schollwoeck2005,
  author     = {Schollw{\"o}ck, U.},
  title      = {The density-matrix renormalization group},
  journal    = {Rev. Mod. Phys.},
  publisher  = {American Physical Society (APS)},
  volume     = {77},
  number     = {1},
  pages      = {259--315},
  year       = {2005},
  month      = apr,
  doi        = {10.1103/RevModPhys.77.259},
  issn       = {1539-0756},
}

@article{Schollwoeck2011,
  author     = {Schollw{\"o}ck, Ulrich},
  title      = {The density-matrix renormalization group in the age of matrix product states},
  journal    = {Ann. Phys.},
  publisher  = {Elsevier BV},
  volume     = {326},
  number     = {1},
  pages      = {96--192},
  year       = {2011},
  month      = jan,
  doi        = {10.1016/j.aop.2010.09.012},
  issn       = {0003-4916},
}

@article{Okamoto1996,
  author     = {Okamoto, Kiyomi and Nomura, Kiyohide},
  title      = {Critical properties of the {XXZ} chain in an external staggered magnetic field},
  journal    = {J. Phys. A: Math. Gen.},
  publisher  = {IOP Publishing},
  volume     = {29},
  number     = {9},
  pages      = {2279--2281},
  year       = {1996},
  month      = may,
  doi        = {10.1088/0305-4470/29/9/036},
  issn       = {1361-6447},
}

@article{Tsukano1998,
  author     = {Tsukano, Masayoshi and Nomura, Kiyohide},
  title      = {{Berezinskii-Kosterlitz-Thouless} transition of spin-1 {XXZ} chains in a staggered magnetic field},
  journal    = {J. Phys. Soc. Jpn.},
  publisher  = {Physical Society of Japan},
  volume     = {67},
  number     = {1},
  pages      = {302--306},
  year       = {1998},
  month      = jan,
  doi        = {10.1143/JPSJ.67.302},
  issn       = {1347-4073},
}

@article{Kuzmenko2009,
  author     = {Kuzmenko, Igor and Essler, Fabian H. L.},
  title      = {Dynamical correlations of the spin-{$1/2$} {Heisenberg} {XXZ} chain in a staggered field},
  journal    = {Phys. Rev. B},
  publisher  = {American Physical Society (APS)},
  volume     = {79},
  number     = {2},
  pages      = {024402},
  year       = {2009},
  month      = jan,
  doi        = {10.1103/PhysRevB.79.024402},
  issn       = {1550-235X},
}

@article{Andraschko2014,
  author     = {Andraschko, F. and Sirker, J.},
  title      = {Dynamical quantum phase transitions and the {Loschmidt} echo: A transfer matrix approach},
  journal    = {Phys. Rev. B},
  publisher  = {American Physical Society (APS)},
  volume     = {89},
  number     = {12},
  pages      = {125120},
  year       = {2014},
  month      = mar,
  doi        = {10.1103/PhysRevB.89.125120},
  issn       = {1550-235X},
}

@article{Rutkevich2018,
  author     = {Rutkevich, Sergei B.},
  title      = {Kink confinement in the antiferromagnetic {XXZ} spin-{$(1/2)$} chain in a weak staggered magnetic field},
  journal    = {EPL},
  publisher  = {IOP Publishing},
  volume     = {121},
  number     = {3},
  pages      = {37001},
  year       = {2018},
  month      = feb,
  doi        = {10.1209/0295-5075/121/37001},
  issn       = {1286-4854},
}

\end{document}